\documentstyle[seceq,epsf,twoside]{ptptex}
\notypesetlogo  
\title{Confirmation of $\sigma$(450--600)-Meson in 
             $\Upsilon^\prime\rightarrow\Upsilon\pi\pi$ $\&$ \\
             Other $\pi\pi$-Production Processes}
\author{%
Muneyuki {\sc Ishida}, Shin {\sc Ishida}$^{*}$  
Toshihiko {\sc Komada}$^{**}$ and\\  Shin-Ichirou {\sc Matsumoto}$^{*}$}
\inst{%
Department of Physics, Tokyo Institute of Technology\\
Tokyo 152-8551, Japan\\
$^{*}$Atomic Energy Research Institute, 
College of Science and Technology\\
Nihon University, Tokyo 101-0062, Japan\\
$^{**}$Department of Engineering Science, Junior College Funabashi Campus,
           Nihon University, Funabashi 274-8501, Japan
}
\recdate{%
\today
}
\abst{%
Applying the effective amplitude, which is evidently consistent with general constraints
from chiral symmetry, the $\pi\pi$ spectra in the relevant processes are analyzed,
leading to a strong evidence for existence of the light $\sigma$ meson.
It is also pointed out that the $\pi\pi$ scattering process, which had been one of the 
main sources for PDG table for these many years, is, in principle, exceptionally
difficult to investigate the property of $\sigma$-meson. 
}

\begin{document}
\maketitle

\setcounter{tocdepth}{4}

\section{Introduction}
In the previous work referred to as I\cite{rf1}, 
we have analyzed systematically the 
$\pi\pi$ production amplitudes in the various excited $\Upsilon$
and $J/\psi$ decay processes, and reproduced successfully the experimental
behaviors:  
The production amplitude ${\cal F}$ is given by a coherent sum of 
$\sigma$ Breit-Wigner amplitude ${\cal F}_\sigma$ and direct
$2\pi$ amplitude ${\cal F}_{2\pi}$ in the VMW method.
\begin{eqnarray}
{\cal F} &=& {\cal F}_\sigma^I + {\cal F}_{2\pi}^I;\ \ 
{\cal F}_\sigma^I =\frac{r_\sigma e^{i\theta_\sigma}}{m_\sigma^2-s-i\sqrt s \Gamma_\sigma (s)},
\ \  {\cal F}_{2\pi}^I=r_{2\pi}e^{i\theta_{2\pi}}  ,
\label{eq1}
\end{eqnarray}
where $r_\sigma$ ($r_{2\pi}$) is
production coupling constant of $\sigma$-state 
($2\pi$-state) and $e^{i\theta_\sigma}$ ($e^{i\theta_{2\pi}}$) 
is a strong phase factor.
Here 
$(r_\sigma ,\theta_\sigma )$ and $(r_{2\pi},\theta_{2\pi})$ are taken to be 
process-dependent, free parameters,
since the $|\sigma\rangle$ state and $|2\pi\rangle$ state are,
from quark physical picture\cite{rfquark}, independent bases 
of $S$-matrix with an independent vertex, in principle.  

However, we must give special attention on the threshold behaviors. 
Because of the  
property of $\pi$ meson as Nambu-Goldstone boson 
in the case of chiral symmetry breaking,  
the $|{\cal F}|^2$ was widely believed to be 
suppressed in the $\pi\pi$ threshold.
In the $\pi\pi$ scattering, the observed spectrum is actually suppressed
near the threshold.
However, in the relevant excited-$\Upsilon$ decay processes 
this threshold suppression is also observed experimentally 
in $\Upsilon (2S\to 1S)$ and $\psi (2S\to 1S)$, 
while, in $\Upsilon (3S)\to\Upsilon (1S)\pi\pi$,
the steep increase from the $\pi\pi$ threshold is observed. 

In the following we 
examine the consistency of our results in I,
especially the threshold behavior of $\Upsilon (3S\to 1S)$, 
with the constraint from chiral symmetry.

\section{Threshold\ Behavior\ of\ Production\ Amplitude\ and\ Chiral\ Symmetry}
\subsection{Effective\ amplitude\ by\ linear\ $\sigma$\ model}
First we consider an 
effective chiral symmteric Lagrangian\cite{rfBC,PLB2}
including $\sigma$ and $\pi$ mesons,
${\cal L}^{(n)}  = 
\xi^{(n)} \Upsilon '_\mu \Upsilon_\mu (\sigma^2+{\bf \pi}^2),
$
where $\Upsilon ' (\Upsilon )$ is the initial (final) 
$b\bar b$ quarkonium.
Through the spontaneous breaking of chiral symmetry, 
$\sigma$ acquires vacuum expectation value,
$\langle\sigma\rangle_0\equiv\sigma_0=f_\pi$, and
${\cal L}^{(n)}$ is rewritten as 
$
{\cal L}^{(n)} =  
\xi^{(n)} \Upsilon '_\mu \Upsilon_\mu 
(f_\pi^2+2f_\pi\sigma'+\sigma'^2+{\bf \pi}^2),
$
producing the coupling terms; 
${\cal L}_\sigma =\xi_\sigma \Upsilon '_\mu \Upsilon_\mu \sigma^\prime \ 
(\xi_\sigma =2f_\pi \xi^{(n)})$,  
${\cal L}_{\pi\pi}=\xi_{2\pi}\Upsilon '_\mu \Upsilon_\mu {\bf \pi}^2\ 
(\xi_{2\pi}=\xi^{(n)})$.
The $\pi\pi$ production amplitude ${\cal F}^{L\sigma M}$ is given by
sum of ${\cal F}_\sigma^{L\sigma M}$ and 
${\cal F}_{2\pi}^{L\sigma M}$, canceling with each other 
in $O(p^0)$ level. 
\begin{eqnarray}
{\cal F}^{L\sigma M} &=& {\cal F}_\sigma^{L\sigma M}  
+ {\cal F}_{2\pi}^{L\sigma M} =
\frac{\xi_\sigma (-2g_{\sigma\pi\pi})}{m_\sigma^2-s}
 + 2\xi^{(n)}
=  
   2\xi^{(n)}\frac{m_\pi^2 -s}{m_\sigma^2-s},\ \ \ \ \ 
\label{eq4}
\end{eqnarray}  
where the relation of SU(2)L$\sigma$M,
$g_{\sigma\pi\pi}=(m_\sigma^2-m_\pi^2)/(2f_\pi )$, is used,
$s=-(p_1+p_2)^2$ ($p_{1,2}$ being the pion momenta), and 
the factor of polarization vectors of initial and final $b\bar b$ quarkonia,
$ \epsilon (P')\cdot \tilde\epsilon (P) $, is omitted. 
The final amplitude takes $O(p^2)$ form, being consistent with the derivative coupling property
of $\pi$ meson. 
In the limit $p_{1\mu}\rightarrow 0_\mu$,
$s\rightarrow m_\pi^2$ and ${\cal F}^{L \sigma M}\rightarrow 0$. 
Thus the amplitude has Adler 0, which
leads to the threshold suppression in conformity
with the observed threshold behavior
in $\Upsilon (2S) \to \Upsilon (1S) \pi\pi$.

\subsection{Effective\ amplitude\ through\ intermediate\ 
           glueball\ states}
Next we consider the effective amplitude through
the intermediate production of the ground-state scalar and tensor 
glueballs.
By using the framework of new classification scheme\cite{rfCLC}, 
the effective interaction is given by
$
{\cal L}^G  \approx  
({\xi^G}/{M'M}) 
\partial_\mu \Upsilon^\prime_\lambda \partial_\nu \Upsilon_\lambda 
(\partial_\mu\sigma\partial_\nu\sigma
+\partial_\mu{\bf \pi}\cdot\partial_\nu {\bf \pi}),
$
where $M'(M)$ is the mass of initial(final) $\Upsilon$.
$\xi^G$ is coupling constant.
It is notable that,
after spontaneous symmetry breaking,
it produces no $\sigma$-amplitude cancelling the direct ${2\pi}$-amplitude.
Then the production amplitude is given by 
\begin{eqnarray}
{\cal F}^G_{2\pi} &=& - \epsilon (P')\cdot \tilde\epsilon (P) 
(\xi^G/M'M) (P'\cdot p_1 P\cdot p_2+P'\cdot p_2 P\cdot p_1)\  ,
\label{eq6}
\end{eqnarray}
where $P'(P)$ is the  momentum of initial(final) $\Upsilon$.

The ${\cal F}^G_{2\pi}$ vanishes
when $p_{1\mu}\to 0_\mu$. Thus, it has Adler 0, satisfying the 
general constraint from chiral symmetry.
However, it does not vanish at $s=m_\pi^2$. 
In the relevant process,  
the Adler limit $p_{1\mu}\to 0_\mu$ corresponds to neglect  
$\Delta M\equiv M'-M$ in comparison with $m_\pi$.
At $s=4m_\pi^2$ the pion four-momenta should be $p_{1\mu}=p_{2\mu}$,
leading to  
$p_{10}=p_{20}\approx \Delta M/2=450$MeV$\gg m_\pi$
in $\Upsilon (3S)\to \Upsilon (1S)$ transition.
Actually ${\cal F}^G_{2\pi}$ can be approximated as 
${\cal F}^G_{2\pi}\approx -2\xi^G p_{10}p_{20}$,
which is almost $s$-independent in all the physical region 
and has no zero close to the threshold.
Correspondingly, there is no suppression near the threshold.

\subsection{Quantitative\ Analysis}
In the actual analysis we modify ${\cal F}^{L\sigma M}$ in Eq.~(\ref{eq4}), 
with inclusion of width of the $\sigma$-meson, 
$\Gamma_\sigma (s)(=g_\sigma^2 p_1(s)/(8\pi s) )$ and of strong phase, 
as
${\cal F}^{\rm phen}_{\sigma +2\pi} = 
(\epsilon (P')\cdot\tilde\epsilon (P))
2 \xi^{(n)} e^{i\theta_\sigma} 
\frac{m_\pi^2-s}{m_\sigma^2-s-i\sqrt{s}\Gamma_\sigma (s)} .$

\begin{figure}[t]
  \epsfxsize=14 cm
  \epsfysize=9 cm
 \centerline{\epsffile{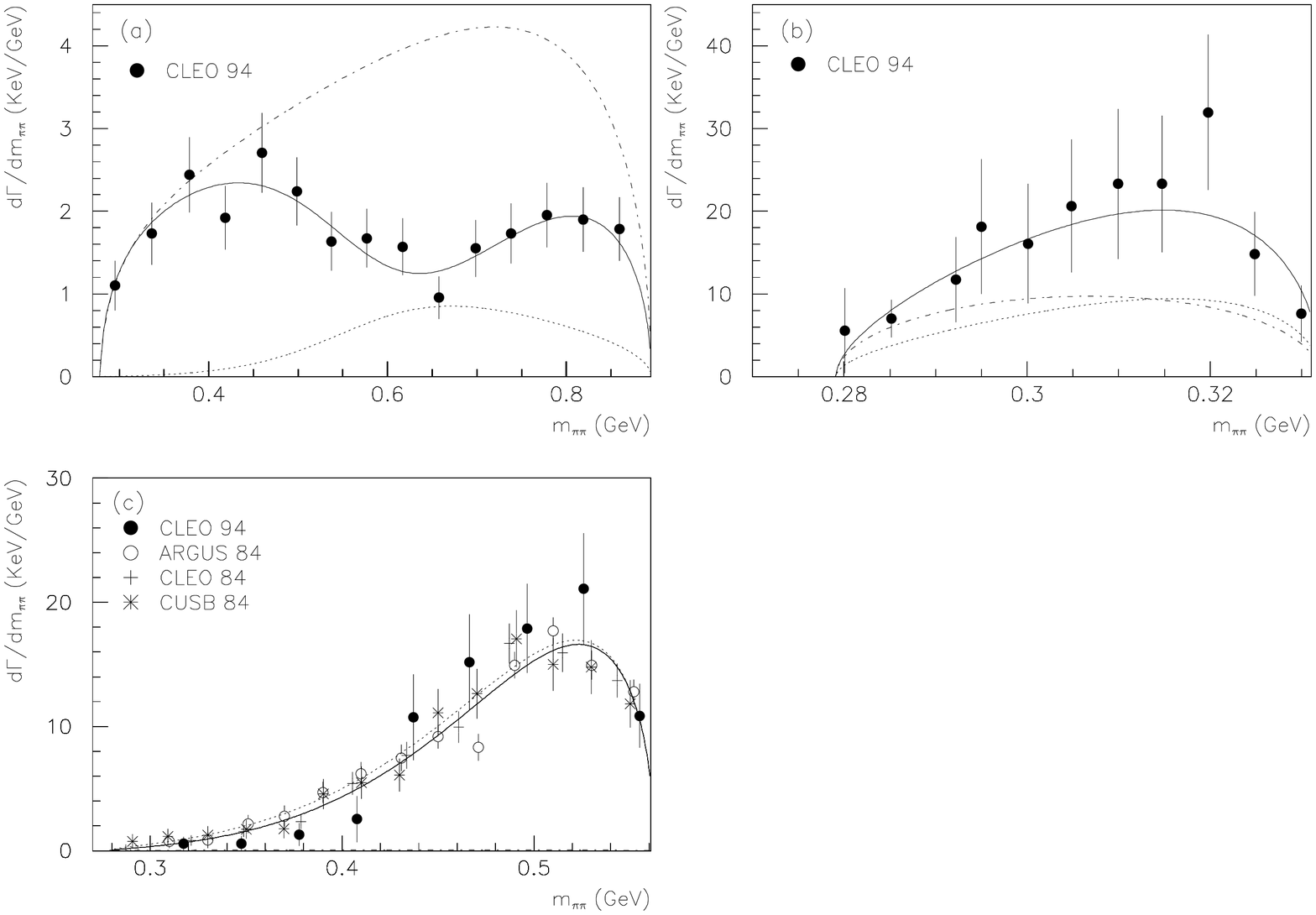}}
 \caption{Fit to the $\pi\pi$ invariant mass spectrum\cite{PLB2}
 by the amplitude 
explicitly consistent with chiral constraint: (a) $\Upsilon (3S)\to \Upsilon (1S)\pi\pi$,
(b) $\Upsilon (3S)\to \Upsilon (2S)\pi\pi$ and (c) $\Upsilon (2S)\to \Upsilon (1S)\pi\pi$.
 The respective contributions from ${\cal F}^{\rm phen}_{\sigma +2\pi}$ and 
from ${\cal F}^G_{2\pi}$ are shown by dotted and dot-dashed lines. 
}
  \label{fig:1}
\end{figure}
The results of the fit 
by using the semi-phenomenological amplitude 
${\cal F}^{\rm phen} \equiv
          {\cal F}^{\rm phen}_{\sigma +2\pi}+{\cal F}^G_{2\pi}$
are depicted in Fig.~1.
The $\pi\pi$ mass spectra in 
$\Upsilon (2S)\to \Upsilon (1S)$, $\Upsilon (3S)\to \Upsilon (1S)$ 
and $\Upsilon (3S)\to \Upsilon (2S)$ transitions are fitted simultaneously 
with common values of 
$m_\sigma$ and $g_\sigma$, leading to the $\sigma$-pole position,
$
\sqrt{s_{\rm pole}} (\approx  m_\sigma -i\Gamma_\sigma /2 )=
(580\stackrel{+79}{\scriptstyle -30})-i(190\stackrel{+107}{\scriptstyle -49}),
$
which is consistent with the ones obtained in our preceding works:
\cite{rf1} and  
the analysis\cite{rfpipi} of $\pi\pi$ scattering phase shift. 
The total $\chi^2$ is $\chi^2 /(N_{\rm data}-N_{\rm param})=54.5/(74-11)=0.87$.
The ${\cal F}^G_{2\pi}$ contribution becomes dominant
in $\Upsilon (3S)\to \Upsilon (1S)$ transition, 
and it explains 
the steep increase from the threshold.

Now we can see our 
treatment in the previous analysis\cite{rf1} is to be consistent with
general chiral constraints:\ \ 
Threshold suppression in $\Upsilon (2S)\rightarrow \Upsilon (1S)$,
required from chiral symmetry, is reproduced phenomenologically by the cancellation
between ${\cal F}_\sigma^I$ and ${\cal F}_{2\pi}^I$ of I (quoted in Eq.~(\ref{eq1})).
Dominant ${\cal F}^G_{2\pi}$-contribution, which is almost $s$-independent
in all physical region, in  $\Upsilon (3S)\rightarrow \Upsilon (1S)$
is reproduced mainly by constant ${\cal F}_{2\pi}^I$ of I .

\subsection{Features\ of\ $\pi\pi$\ production\ amplitudes}
The $\pi\pi$ production processes have generally 
much the larger energy release $\Delta E$ 
than $m_\pi$. Thus, the momentum of emitted pion becomes large, and the
derivative type amplitude, as Eq.(\ref{eq6}), 
may play an important role. The Adler 0 generally does not imply 
the suppression at the small $s$ region, and the $\pi\pi$ spectrum 
shows possibly steep increase from threshold.
There is no mechanism 
cancelling the effect of $\sigma$ production, and 
the direct $\sigma$-peak structure is expected to be observed. 
Actually,\footnote{
Concerning the $pp$-central collision experiment 
$pp\to pp (\pi^0\pi^0)$ by GAMS\cite{rfpp}, which also seems to suggest
the $\sigma$-existence, see the comment in the talk by T. Komada, 
this conference\cite{rf1}.
}  
in $J/\psi \to \omega\pi\pi$ decay\cite{rfJpsi,rf8},
$p\bar p \to 3\pi^0$,\cite{rfppbar} and 
$D^-\rightarrow\pi^-\pi^+\pi^+$\cite{gobel},
where $\Delta E$ is very large, 
the $\sigma$-peak structure is observed directly. 
In the relevant $\Upsilon$ decay processes
$\Delta E$ in ${\Upsilon (3S\to 1S)}$ is the largest, and
only this process shows the steep increase from threshold.

\begin{figure}[t]
  \epsfxsize=11 cm
  \epsfysize=7 cm
 \centerline{\epsffile{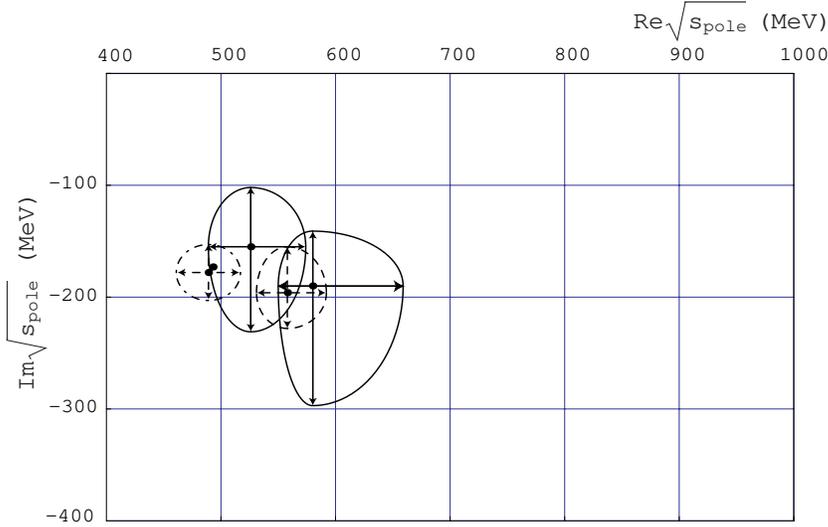}}
 \caption{Pole positions $\sqrt{s_{\rm pole}}\approx m_\sigma -i\Gamma_\sigma /2$ of 
$\sigma$ and their error regions, 
obtained from the various $\pi\pi$-production processes:
$\Upsilon$ decays \cite{rf1} and \cite{PLB2},
$p\bar p\rightarrow 3\pi^0$\cite{rfppbar}, 
$D^-\rightarrow\pi^-\pi^+\pi^+$\cite{gobel} and 
$J/\psi\rightarrow\omega\pi\pi$\cite{rf8}
are shown by the points with crosses. Respective 
regions (except for $J/\psi$ decay)
are surrounded by solid lines, dashed line and dot-dashed line.
}
  \label{fig:2}
\end{figure}

On the contrary in the $\pi\pi$ scattering process, 
the pion momentum itself becomes small near threshold region,
and the spectrum close to threshold is suppressed.
Because of this chiral constraints, the $\sigma$-amplitude
must be cancelled out by the non-resonant repulsive amplitude,
and the direct $\sigma$-peak 
cannot be seen in $\pi\pi$-scattering, in principle. 


\section{Concluding\ Remarks}
Through the above discussion it proves not adequate to 
determine the pole-position of $\sigma$
only through the $\pi\pi$-scattering.
However,  
the present estimation of the $\sigma$ pole position in PDG table
has been done mainly through the analyses of $\pi\pi$ scattering.
This is one of the reasons why the present label
of $\sigma$, ``$f_0(400$--1200) or $\sigma$", is largely uncertain 
and the $\sigma$ is still regarded as controversial.   

The $\sigma$-pole position should be determined regarding 
the $\pi\pi$-production processes with large energy release, 
which are free from the chiral constraints. 
The pole positions of $\sigma$ obtained through various 
$\pi\pi$-production processes are shown in Fig.~2.
They are almost process-independent, and in the 
range $m_\sigma$=450--600MeV.
Basing on these results we conclude that 
{\it the existence of $\sigma (450$--600) is confirmed, 
and the present label of PDG table
should be corrected accordingly.}

In this connection it is notable that, in this conference, a firm experimental evidence
for existence of the $I=1/2$ scalar $\kappa$-meson,
to be the flavor partner of $\sigma$-meson, is reported. 
This $\kappa$ is observed\cite{gobel} as a peak in the $K\pi$-mass spectra 
in the decay process $D^-\rightarrow K^-\pi^+\pi^+$.
This process is actually a $K\pi$-production process with the above mentioned 
large energy release, being free from the chiral constraints.
The criticism\cite{nokappa} that no such pole is observed in 
$K\pi$-scattering process is disregarding the property of $K$ meson as 
Nambu-Goldstone boson in the case of chiral symmetry breaking. 
This is the same situation as the direct $\sigma$-peak being not 
observed in $\pi\pi$-scattering.
Actually by taking into account this constraint from chiral symmetry,
the $K\pi$-scattering data were shown to be consistent with the exsitence 
of $\kappa$\cite{kappa}.



\end{document}